\documentclass[12pt,showpacs,amsmath,amssymb]{revtex4}

\usepackage{setspace}

\usepackage{graphicx}
\usepackage{epsfig}
\usepackage{amssymb}
\usepackage{graphicx}
\usepackage{dcolumn}
\usepackage{bm}
\newcommand{\ba}{\begin{array}}
\newcommand{\ea}{\end{array}}
\def\br{\begin{eqnarray}}
\def\er{\end{eqnarray}}
\def\be{\begin{equation}}
\def\ee{\end{equation}}

\def\({\left(}
\def\){\right)}

\begin{document}


\title{Leading Pomeron Contributions and the TOTEM Data at 13 TeV}

\author{M. Broilo$^{1}$\footnote{mateus.broilo@ufrgs.br}, E.G.S. Luna$^{1}$\footnote{luna@if.ufrgs.br} and M.J. Menon$^{2}$\footnote{menon@ifi.unicamp.br}}
\affiliation{$^{1}$Instituto de F\'{\i}sica,  Universidade Federal do Rio Grande do Sul \\ CEP 91501-970, Porto Alegre -- RS, Brazil \\ 
$^{2}$Instituto de F\'{\i}sica Gleb Wataghin, Universidade Estadual de Campinas \\ CEP 13083-859, Campinas -- SP, Brazil}  
 

\begin{abstract}

The recent data collected by the TOTEM Collaboration on $\sigma_{tot}$ and $\rho$ at 13 TeV have shown agreement with a leading Odderon contribution at the highest energies, as demonstrated in the very recent analysis by Martynov and Nicolescu (MN). In order to investigate the same dataset by means of Pomeron dominance, we introduce a general class of forward scattering amplitude, with leading contributions even under crossing, associated with simples, double and triple poles in the complex angular-momentum plane. For the lower energy region, we consider the usual nondegenerated Regge trajectories with even and odd symmetry. The analytic connection between $\sigma_{tot}$ and $\rho$ is obtained by means of dispersion relations and we carry out fits to $pp$ and $\bar{p}p$ data in the interval $\sqrt{s}=5$ GeV -- 13 TeV; following MN we consider only the TOTEM data at the LHC energy region. From the fits, we conclude that the general analytic model, as well as some particular cases representing standard parameterizations, are not able to satisfactorily describe the $\sigma_{tot}$ and $\rho$ data at 13 TeV. Further analyses in course and some
perspectives are outlined.
\end{abstract}

\pacs{13.85.-t, 13.85.Lg, 11.10.Jj, 13.85.Dz}

\maketitle

\section{Introduction}

The recent TOTEM results at 13 TeV, $\sigma_{tot}=110.6 \pm 3.4\ \mathrm{mb}$, $\rho=0.10 \pm 0.01$  and $0.09 \pm 0.01$
\cite{totem}, suggest the dominance of the Odderon contribution within the LHC energy region, 
disfavoring the even under crossing Pomeron dominance. 
In the recent analysis by Martynov and Nicolescu (MN), these two measurements are quite well described with the Odderon
in its maximal form  \cite{martynico}. The analysis treats $pp$ and $\bar{p}p$ scattering in the interval
5 GeV -- 13 TeV, including only the TOTEM data at the LHC energy region.
Here, we analyze the same dataset through a most general analytic form 
related to the Pomeron dominance and some particular cases too.
After introducing the analytic models, we present the fit results,
followed by our conclusions and an outlining of some results and analyses in course.

\section{General Analytic Model}
In the Regge-Gribov theory, simple, double, and triple poles in the complex angular momentum
plane are associated with power, logarithmic, and logarithmic squared functions for the total
cross section in terms of the energy (see Appendix B in \cite{fms17b} for a recent short review
and references). In this context, we consider a general parameterization for $\sigma_{tot}(s)$
consisting of two Reggeons (even and odd under crossing) and four (even) Pomeron
contributions:

\begin{eqnarray}
\sigma_{\mathrm{tot}}(s) =  a_1 \left[\frac{s}{s_0}\right]^{-b_1} \!\!\!\!\!\!+ \tau a_2 \left[\frac{s}{s_0}\right]^{-b_2}
\!\!\!\!\!\!+ A + B \left[\frac{s}{s_0}\right]^{\epsilon}+ C \ln\left(\frac{s}{s_0}\right) + D \ln^2\left(\frac{s}{s_0}\right).
\label{stg}
\end{eqnarray}
The analytic results for $\rho(s)$ have been obtained by means of singly subtracted Derivative Dispersion Relations \cite{am04}, taking into account an effective subtraction constant $K$ \cite{fms17b}:

\begin{eqnarray}
\!\!\!\!\!\!\rho(s) &=& \frac{1}{\sigma_{\mathrm{tot}}(s)}
\left\{ \frac{K}{s}  
- a_1\,\tan \left( \frac{\pi\, b_1}{2}\right) \left[\frac{s}{s_0}\right]^{-b_1} \!\!\!\!\!\!+
\tau \, a_2\, \cot \left(\frac{\pi\, b_2}{2}\right) \left[\frac{s}{s_0}\right]^{-b_2} \right. \nonumber \\
&+& \left. B\,\tan \left( \frac{\pi\, \epsilon}{2}\right) \left[\frac{s}{s_0}\right]^{\epsilon} +
\frac{\pi C}{2} + \pi D \ln\left(\frac{s}{s_0}\right) \right\}.
\label{rhog}
\end{eqnarray}
These expressions bring enclosed analytic structures \textit{similar} to those appearing in some well-known models,
for example, Donnachie and Landshoff ($A=C=D=0$) \cite{dl}, Block and Halzen ($B=0, \epsilon=0$) \cite{bh},
COMPETE and PDG parameterizations ($B=C=0, \epsilon=0$) \cite{compete,pdg16}.

Here, following \cite{fms17b,fms17a}, we consider the energy scale fixed at
the physical threshold for scattering states,
$s_0 = 4m_p^2 \approx $ 3.521 GeV$^2$,
where $m_p$ is the proton mass. With this assumption, we can identify two particular
and independent cases, to be denoted by




\vspace{0.3cm}

{\bf Model I}: $A=C=D=0$ 

\vspace{0.3cm}

{\bf Model II}:  $B=C=0, \epsilon=0$

\vspace{0.3cm}

\noindent
In the General Model,  Eqs. (\ref{stg}) and (\ref{rhog}), there are 10 free parameters: $a_1$, $b_1$, $a_2$, $b_2$, $A$, $B$, $\epsilon$, $C$, $D$ and $K$, which are determined by fits to the experimental data on $\sigma_{tot}$ and $\rho$ from $pp$ and $\bar{p}p$ elastic scattering in the interval of 5 GeV -- 13 TeV.

\section{Fits and Results}

The data below 7 TeV have been collected from the PDG database \cite{pdg16} without any kind of data selection. At 7, 8, and 13 TeV, the dataset includes only the TOTEM data (11 points). In each case (Models I, II and the General Model), we consider two variants: either 
$K$ as a free fit parameter or $K = 0$ (fixed). 

The fits were performed with the TMinuit package and using the default MINUIT error analysis
\cite{minuit}.
As convergence criterium we consider only minimization results which imply positive-definite covariance matrices. As tests of goodness-of-fit we shall consider the chi-square per degree of freedom, $\chi^2/\nu$, and the integrated probability,
$P(\chi^2)$.

The data reductions with the General Model did not comply with the above convergence requirements and thus can not be regarded as a possible solution. This may be due to an excessive number of free parameters. On the other hand, in particular cases as given by Models I and II, the convergence criteria were reached. The statistical information on these fits are displayed in Table 1 and the corresponding curves are depicted in Figure 1.

\begin{table*}[!ht]
\caption{Fit results to $\sigma_{tot}$ and $\rho$ data from
$pp$ and $\bar{p}p$ scattering through Models I and II
and their variants: $K$ as a free fit parameter and $K=0$ (fixed).}
\begin{tabular}{c@{\quad}c@{\quad}c@{\quad}c@{\quad}c@{\quad}}
\hline \hline
Model:     & I & I & II & II    \\[0.05ex]
\hline
& & & &\\[-0.2cm]
$a_{1}$ (mb)            & 41.40\,$\pm$\,0.61      & 40.31\,$\pm$\,0.50   &
32.18\,$\pm$\,0.65  & 31.34\,$\pm$\,0.47 \\[0.05ex]
$b_{1}$                 & 0.3776\,$\pm$\,0.0010    & 0.3586\,$\pm$\,0.0084 &
0.392\,$\pm$\,0.017 & 0.3637\,$\pm$\,0.0078 \\[0.05ex]
$a_{2}$ (mb)            & 17.02\,$\pm$\,0.72      & 17.50\,$\pm$\,0.73   &
16.98\,$\pm$\,0.72   & 17.32\,$\pm$\,0.72 \\[0.05ex]
$b_{2}$                 & 0.545\,$\pm$\,0.012    & 0.557\,$\pm$\,0.013 &
0.545\,$\pm$\,0.013 & 0.554\,$\pm$\,0.013 \\[0.05ex]
$A$ (mb)                & -                      & -                   &
29.59\,$\pm$\,0.41  & 28.95\,$\pm$\,0.21 \\[0.05ex]
$B$ (mb)                & 21.62\,$\pm$\,0.26      & 21.20\,$\pm$\,0.24   & -          
        & - \\[0.05ex]
$\epsilon$              & 0.0914\,$\pm$\,0.0011    & 0.0929\,$\pm$\,0.0010 & -          
        & - \\[0.05ex]
$D$ (mb)                & -                      & -                   &
0.2512\,$\pm$\,0.0037 & 0.2549\,$\pm$\,0.0024 \\[0.05ex]
$K$ (mb\,GeV$^{2}$) & 69\,$\pm$\,17      & 0 (fixed)           &
55 \,$\pm$\,18   & 0 (fixed)\\[0.05ex]
\hline
& & & & \\[-0.2cm]
$\nu$                   & 248                    &  249                & 248        
        & 249  \\[0.05ex]
$\chi^2/\nu$            & 1.273                  &  1.339              & 1.192      
        & 1.229   \\[0.05ex]
$P(\chi^2)$             & 2.3 $\times$ 10$^{-3}$        &  2.8 $\times$ 10$^{-4}$    & 2.0 $\times$
10$^{-2}$     & 7.9 $\times$ 10$^{-3}$ \\[0.05ex]
\hline \hline 
\end{tabular}
\label{t1}
\end{table*}

\begin{figure}[!ht]
\centering
\includegraphics[width=13cm,height=12cm]{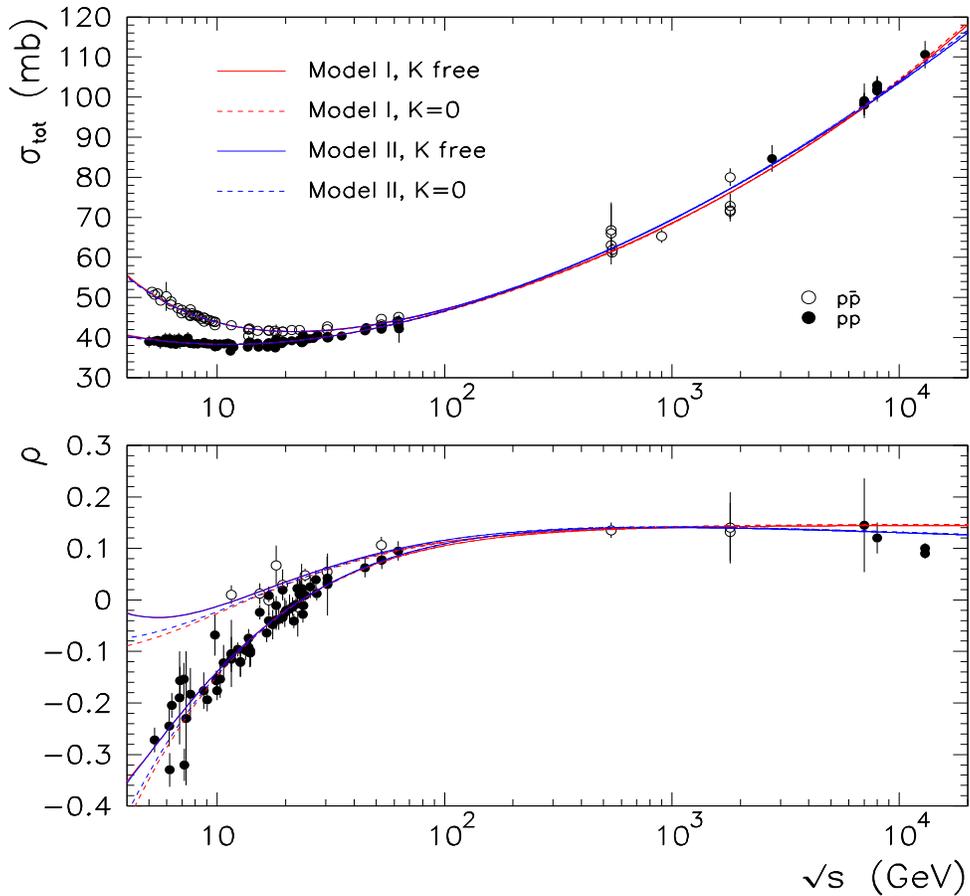}
\label{fig1}
\caption{Fit results to the total cross section (upper) and $\rho$ ratio (lower) of $pp$ 
and $\bar{p}p$ scattering data through Model I (red) and Model II (blue).}
\end{figure}

\section{Conclusions and Perspectives}
From Figure 1 we see that \textit{at 13 TeV}, the closest curve to the central value of $\sigma_{tot}$ datum corresponds to the most distant one from the  $\rho$ data (Model I). Both model results cross the lower error bar of $\sigma_{tot}$, but lie well above the upper error bar of the  $\rho$ data.
From Table I, the integrated probability is rather small. 
These results do not depend on the variant considered, namely $K$ as a free fit parameter or $K=0$ (fixed).
We conclude that, within the dataset considered, none of the models present simultaneous agreement with the $\sigma_{tot}$  and $\rho$ data at 13 TeV.

Following MN, we did not include the ATLAS measurements on $\sigma_{tot}$ at 7 and 8 TeV.
This, however, might not be a correct exclusion, mainly due to the discrepancy between the TOTEM and ATLAS data
\cite{fms17a}. Moreover, beyond the General Model, we have considered only two particular cases,
without tests with all possible variants. Most importantly, as done in the MN analysis, we did not evaluate
the uncertainties regions in the fit results (curves). 

We are presently investigating all these aspects, including the ATLAS data and uncertainty regions 
through error propagation from the 
fit parameters with confidence level of 90\%. We anticipate that one variant seems
not to be excluded by the bulk of experimental data presently available at the LHC energy
region \cite{blm}.

\section*{Acknowledgments}

This research was partially supported by the Conselho Nacional de Desenvolvimento Cient\'{\i}fico e Tecnol\'ogico (CNPq) and by the Funda\c{c}\~ao de Amparo \`a Pesquisa do Estado do Rio Grande do Sul (FAPERGS). E.G.S.L acknowledge the financial support from the Rede Nacional de Altas Energias (RENAFAE).

\end{document}